\documentclass[twocolumn,eqsecnum,showkeys,showpacs,nofootinbib,aps,epsfig]{revtex4}
\renewcommand{\theequation}{\arabic{equation}}
\usepackage{graphicx}
\def\bea{\begin{eqnarray}}
\def\eea{\end{eqnarray}}

\newcommand{\nn}{\nonumber}
\newcommand{\na}{\nabla}
\def\beq{\begin{equation}}
\def\eeq{\end{equation}}

\begin{document}
\input epsf
\title{Hydrodynamics and global structure of rotating Schwarzschild black holes}
\author{Soon-Tae Hong}
\email{soonhong@ewha.ac.kr} \affiliation{Department of Science
Education, Ewha Womans University, Seoul 120-750 Korea}
\author{Sung-Won Kim}
\email{sungwon@ewha.ac.kr} \affiliation{Department of Science
Education, Ewha Womans University, Seoul 120-750 Korea}
\affiliation{Asia Pacific Center for Theoretical Physics, Pohang
790-784 Korea}
\date{\today}%
\begin{abstract}
Exploiting a rotating Schwarzschild black hole metric, we study
hydrodynamic properties of perfect fluid whirling inward toward
the black holes along a conical surface.  On the equatorial plane
of the rotating Schwarzschild black hole, we derive radial
equations of motion with effective potentials and the Euler
equation for steady state axisymmetric fuid. Moreover, numerical
analysis is performed to figure out effective potentials of
particles on the rotating Schwarzschild manifolds in terms of
angular velocity, total energy and angular momentum per unit rest
mass.  Higher dimensional global embeddings are also constructed
inside and outside the event horizons of the rotating
Schwarzschild black holes.
\end{abstract}
 \maketitle

\section{Introduction}
\setcounter{equation}{0}
\renewcommand{\theequation}{\arabic{section}.\arabic{equation}}

The physics of a rotating charged sphere has long attracted the
attention of physicists~\cite{schiff,blackett,ruffini}.  From the
experimental viewpoint, the pulsars have given concrete evidence
for the existence of rotating magnetized collapsed objects.  From
the theoretical viewpoint, the existing exact solutions of
Einstein equations have shown that the most general stationary
solution, which is asymptotically flat with a regular horizon for
a fully collapsed object, has to be rotating and endowed with a
net charge.  It is well known that the Kerr~\cite{kerr} family of
solutions of the Einstein and Einstein-Maxwell equations is the
general class of solutions which could represent the field of a
rotating neutral or electrically charged sphere in asymptotically
flat space.  In the extended manifolds, all geodesics which do not
reach the central ring singularities of the Kerr black hole are
shown to be complete, and also those null or timelike geodesics
which do reach the singularities are entirely confined to the
equator~\cite{carter}.  Moreover, the Kerr metric has the region
called the ergosphere where the asymptotic time translation
Killing field becomes spacelike.  In the ergosphere, all observers
are forced to rotate in the direction of the rotation of the black
hole. Recently, the rotating Schwarzschild wormhole metric was
proposed to investigate classes of geodesics falling through it
which do not encounter any energy condition violating
matter~\cite{teo98}.

On the other hand, a familiar feature of exact solutions to the
field equations of general relativity is the presence of
singularities. As novel ways of removing the coordinate
singularities, the higher dimensional global flat embeddings of
the black hole solutions are subjects of great interest both to
mathematicians and to physicists.  In differential geometry, it
has been well-known that four dimensional Schwarzschild
metric~\cite{sch} is not embedded in $R^{5}$~\cite{spivak75}.
Recently, (5+1) dimensional global embedding Minkowski space
(GEMS) structure for the Schwarzschild black hole has been
obtained~\cite{deser97} to investigate a thermal Hawking effect on
a curved manifold~\cite{hawk75} associated with an Unruh
effect~\cite{unr} in these higher dimensional spacetime.  In (3+1)
dimensions, the global flat embeddings inside and outside of event
horizons of Schwarzschild and Reissner-Nordstr\"{o}m black holes,
have been constructed and on these overall patches of the curved
manifolds four accelerations and Hawking temperatures have been
evaluated by introducing relevant Killing
vectors~\cite{honggrg04}. Recently, the GEMS scheme has been
applied to stationary motions in spherically symmetric
spacetime~\cite{chen04}, and the Banados-Teitelboim-Zanelli black
hole~\cite{btz} has been embedded in (3+2) dimensions to
investigate the SO(3,2) global and Sp(2) local
symmetries~\cite{hongplb04}.

In this paper, we take an ansatz for a rotating Schwarzschild
black hole to investigate hydrodynamic properties of the perfect
fluid spiraling inward toward the black holes along a conical
surface.  On the rotating Schwarzschild black hole manifolds we
then construct higher dimensional global embeddings inside and
outside the event horizons of the black holes. We also perform
numerical evaluations of effective potentials of particles on the
equatorial planes of the rotating Schwarzschild black holes in
terms of angular velocity, total energy and angular momentum per
unit rest mass.

This paper is organized as follows. In section II we study the
rotating Schwarzschild black hole with constant angular velocity,
and in section III we investigate that with angular velocity
proportional to $1/r^{3}$. Section IV includes summary and
discussions.

\section{Rotating Schwarzschild black hole with constant $\Omega$}
\setcounter{equation}{0}
\renewcommand{\theequation}{\arabic{section}.\arabic{equation}}

We consider the rotating Schwarzschild black hole with a constant
angular velocity $\Omega$ whose 4-metric is described as \beq
ds^{2}=-N^{2}dt^{2}+N^{-2}dr^2+r^{2}d\theta^{2}+r^{2}\sin^{2}\theta(d\phi-\Omega
dt)^{2}. \label{4metric1}\eeq Here in the units $G=c=1$ the lapse
function $N^{2}$ is defined as \beq N^{2}=1-\frac{2m}{r}= \frac{r-
r_{H}}{r}. \label{schlapse} \eeq  The event horizon $r_{H}$ is
located at the pole of $g_{rr}$, namely at the root of $N^{2}$ to
yield $r_{H}=2m$ as in the static Schwarzschild black hole case.
The four velocity is given by \beq u^{a}=\frac{d x^{a}}{d
\tau},\label{ua}\eeq where we can choose $\tau$ to be the proper
time (affine parameter) for timelike (null) geodesics. From the
equation of motion of a test particle in the rotating
Schwarzschild manifold, the particle initially at rest at infinity
spiral inward toward the black hole along a conical surface of
constant $\theta=\theta_{\infty}$ where $\theta_{\infty}$ is the
polar angle at infinity.  For a fluid which is at rest at infinity
and approaches supersonically to the black hole, one may take the
approximation to simplify the hydrodynamical equations \beq
u^{\theta}=\frac{d\theta}{d\tau}\approx 0. \eeq

As in the Schwarzschild black hole since the coordinates $t$ and
$\phi$ are cyclic we have the timelike Killing field $\xi^{a}$ and
the axial Killing field $\psi^{a}$. Corresponding to the Killing
fields $\xi^{a}$ and $\psi^{a}$ we can then find the conserved
energy $E$ and the angular momentum $L$ per unit rest mass for
geodesics given as follows \bea
E&=&-g_{ab}\xi^{a}u^{b}\nn\\&=&\left(\frac{r-r_{H}}{r}-\Omega^{2}r^{2}\sin^{2}\theta\right)
~u^{t}+\Omega r^{2}\sin^{2}\theta~u^{\phi},\nn\\
L&=&g_{ab}\psi^{a}u^{b}=-\Omega
r^{2}\sin^{2}\theta~u^{t}+r^{2}\sin^{2}\theta~u^{\phi},
\label{eandl} \eea where $u^{a}$ are four velocity of the locally
nonrotating observers defined by (\ref{ua}). Moreover, we can
introduce a new conserved parameter $\kappa$ defined as \beq
\kappa=-g_{ab}u^{a}u^{b}\label{kappaeq}\eeq whose values are given
by $\kappa=1$ for timelike geodesics and $\kappa=0$ for null
geodesics.

In the case of geodesics on the equatorial plane $\phi=\pi/2$,
$u^{t}$ and $u^{\phi}$ are given in terms of $E$ and $L$ as
follows \bea
u^{t}&=&\frac{r}{r-r_{H}}E-\frac{\Omega r}{r-r_{H}}L\nn\\
u^{\phi}&=&\frac{\Omega r}{r-r_{H}}E
+\left(\frac{1}{r^{2}}-\frac{\Omega^{2}r}{r-r_{H}}\right)L,
\label{tphidot}\eea which are substituted into (\ref{kappaeq}) to
yield the radial equation for the particle on the equatorial plane
\beq
\frac{1}{2}E^{2}=\frac{1}{2}u^{r}u^{r}+V(r,E,L),\label{eomr}\eeq
with the effective potential \beq
V=-\frac{r_{H}}{2r}\kappa+\frac{L^{2}}{2r^{2}}+\frac{1}{2}\kappa-\frac{L}{r^{3}}
\left[\left(\frac{r_{H}}{2}+\frac{\Omega^{2}r^{3}}{2}\right)L-\Omega
r^{3}E\right]. \label{effpot}\eeq Here the first and second terms
denote the Newtonian and centrifugal barrier terms respectively,
which are attainable from Newtonian mechanics, while the other
terms are general relativistic corrections, including the black
hole rotating effects with the parameter $\Omega$.  If $E<1$ the
orbit of the particle is bound so that it cannot reach infinity,
while if $E>1$ the orbit is unbound except for a measure-zero set
of unstable orbits~\cite{wilkins}.

\begin{figure}
\begin{center}
\includegraphics[width=10cm]{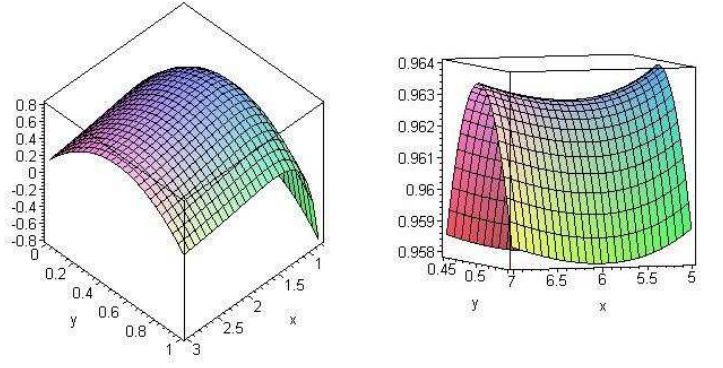}\\
\end{center}
\vskip -.5cm \caption[fig1] {Effective potentials $V(x,y)$ with
$x=r/r_{H}$ and $y=\Omega r_{H}$ for null and timelike geodesics
of particles with $E=1$ and $L=2r_{H}$.} \label{fig1}
\end{figure}

For the null geodesics with $\kappa=0$, we find the only extremum
of the effective potential to be a maximum occurring at
$r=3r_{H}/2$ as in the case of static Schwarzschild black hole.
The effective potential $V(x,y)$ for the particles with the total
energy per unit rest mass $E=1$ and angular momentum per unit rest
mass $L=2r_{H}$ is shown in the left graph in Fig. 1 where $x$ and
$y$ denote the dimensionless variables $x=r/r_{H}$ and $y=\Omega
r_{H}$, respectively.  Here note that we have a maximal effective
potential at the position $(x,y)=(3/2,1/2)$.

Next, we consider the timelike geodesics. The effective potential
(\ref{effpot}) with $\kappa=1$ now should fulfil the condition
\beq \frac{dV}{dr}=0,\label{v0}\eeq to yield the radii of the
stable and unstable bound orbits on the equatorial plane for
$L^{2}>3r_{H}^{2}$ \beq r_{s,us}=\frac{L^{2}\pm
L(L^{2}-3r_{H}^{2})^{1/2}}{r_{H}}\eeq where the upper (lower) sign
refers to the stable (unstable) orbit.  In particular, for the
case of $L\gg r_{H}$, we can find $r_{s}\approx 2L^{2}/r_{H}$
corresponding to the Newtonian radius of circular orbits of
particles with angular momentum per mass $L$ orbiting a spherical
body of mass $m$. The energy per unit mass of the particle in the
circular orbit of the radius $r=r_{s}$ is the value of the
effective potential $V$ at that radius \beq
\frac{1}{2}E^{2}=V(r),\eeq which yields, together with (\ref{v0}),
the energy per unit mass \beq
E_{s}=\frac{2^{1/2}(r_{s}-r_{H})}{r_{s}^{1/2}(2r_{s}-3r_{H})^{1/2}}
+\frac{\Omega r_{H}^{1/2}r_{s}}{(2r_{s}-3r_{H})^{1/2}}.\eeq Note
that we have an term proportional to $\Omega$ additional to the
static Schwarzschild black hole result.  If a particle is
displaced slightly from the equilibrium radius $r_{s}$ of the
stable circular orbit, the particle will oscillate in the radius
about $r_{s}$ to execute simple harmonic motion with frequency
$\omega_{r}$ given by \beq
\omega_{r}=\frac{r_{H}^{1/2}(r_{s}-3r_{H})^{1/2}}{r_{s}^{3/2}
(2r_{s}-3r_{H})^{1/2}}.\label{omegar} \eeq On the other hand, the
angular frequency $\omega_{\phi}$ for the circular orbit is found
to be \beq \omega_{\phi}=\frac{r_{H}^{1/2}+2^{1/2}\Omega
r_{s}^{3/2}}{r_{s}(2r_{s}-3r_{H})^{1/2}}. \eeq  Here note that the
frequency $\omega_{r}$ is the same as that in the static
Schwarzschild black hole case, while the angular frequency
$\omega_{\phi}$ is enhanced by the additional term proportional to
$\Omega$.  For the timelike geodesics of the particles with the
total energy per unit rest mass $E=1$ and angular momentum per
unit rest mass $L=2r_{H}$, the effective potential $V(x,y)$ has a
shape similar to that for the null geodesic case shown in Fig. 1.
However, along the curve $y=1/2$ we have maximal potential values
to yield a maximal effective potential at the position
$(x,y)=(2,1/2)$ and a saddle point of the effective potential at
the position $(x,y)=(6,1/2)$ as shown in the right graph in Fig.
1.

Now, we consider the global aspects of the rotating Schwarzschild
black hole.  After some algebra, for the rotating Schwarzschild
black hole in the region $r\ge r_{H}$ we can obtain the (5+1) GEMS
structure \beq
ds^{2}=-(dz^{0})^{2}+(dz^{1})^{2}+(dz^{2})^{2}+(dz^{3})^{2}+(dz^{4})^{2}+(dz^{5})^{2}
\label{ds1} \eeq with the coordinate transformations \bea
z^{0}&=&k_{H}^{-1}\left(\frac{r-r_{H}}{r}\right)^{1/2}\sinh k_{H}t,\nn\\
z^{1}&=&k_{H}^{-1}\left(\frac{r-r_{H}}{r}\right)^{1/2}\cosh k_{H}t,\nn\\
z^{2}&=&r\sin\theta\cos(\phi-\Omega t),\nn\\
z^{3}&=&r\sin\theta\sin(\phi-\Omega t),\nn\\
z^{4}&=&r\cos\theta,\nn\\
z^{5}&=&\int dr
\left(\frac{r_{H}(r^{2}+r_{H}r+r_{H}^{2})}{r^{3}}\right)^{1/2}
\equiv f(r,r_{H}),\nn\\ \label{gems6} \eea where the surface
gravity $k_{H}$ is given by \beq k_{H}=\frac{1}{2r_{H}}.
\label{sg} \eeq Here we recall that the static Schwarzschild is
embedded in (5+1) dimensions with the GEMS structure (\ref{ds1})
and the coordinate transformations for $r\geq
r_{H}$~\cite{deser97} \bea
z^{0}&=&k_{H}^{-1}\left(\frac{r-r_{H}}{r}\right)^{1/2}\sinh
k_{H}t,
\nonumber \\
z^{1}&=&k_{H}^{-1}\left(\frac{r-r_{H}}{r}\right)^{1/2}\cosh
k_{H}t,
\nonumber \\
z^{2}&=&r\sin\theta\cos\phi,
\nonumber\\
z^{3}&=&r\sin\theta\sin\phi,
\nonumber\\
z^{4}&=&r\cos\theta\nn\\
z^{5}&=&f(r), \label{sch6} \eea where $f(r)$ is read off from
(\ref{gems6}). Here one can readily check that in the limit
$\Omega\rightarrow 0$ the embedding (\ref{gems6}) reduces to that
of the Schwarzschild case (\ref{sch6}).  In order to investigate
the region $r\le r_{H}$ we rewrite the rotating Schwarzschild
black hole four-metric (\ref{4metric1}) as \beq
ds^{2}=\bar{N}^{2}dt^{2}-\bar{N}^{-2}dr^2+r^{2}d\theta^{2}+r^{2}\sin^{2}\theta(d\phi-\Omega
dt)^{2}, \eeq in terms of the positive definite lapse function
inside the event horizon $r_{H}$ \beq
\bar{N}^{2}=-1+\frac{2M}{r}=\frac{r_{H}-r}{r} \label{nbar}\eeq to
yield the (5+1) GEMS structure (\ref{ds1}) with the coordinate
transformation \bea
z^{0}&=&k_{H}^{-1}\left(\frac{r_{H}-r}{r}\right)^{1/2}\cosh k_{H}t,\nn\\
z^{1}&=&k_{H}^{-1}\left(\frac{r_{H}-r}{r}\right)^{1/2}\sinh k_{H}t,\nn\\
z^{5}&=&f(r), \label{gems62} \eea
with $(z^{2},z^{3},z^{4})$ in (\ref{gems6}).


\section{Rotating Schwarzschild black hole with $\Omega=2a/r^{3}$}
\setcounter{equation}{0}
\renewcommand{\theequation}{\arabic{section}.\arabic{equation}}

Now, we consider the rotating Schwarzschild black hole with the
angular velocity of the form $\Omega=2a/r^{3}$~\cite{teo98} by
introducing 4-metric \beq
ds^{2}=-N^{2}dt^{2}+N^{-2}dr^2+r^{2}d\theta^{2}+r^{2}\sin^{2}\theta\left(d\phi-\frac{2a}{r^{3}}
dt\right)^{2}, \label{4metric2}\eeq where $a$ is the angular
momentum of the black hole.  As in the rotating Schwarzschild
black hole case with constant $\Omega$ in the previous section, we
have the timelike Killing field $\xi^{a}$ and the axial Killing
field $\psi^{a}$ to yield the conserved energy $E$ and the angular
momentum $L$ per unit rest mass for geodesics
\bea
E&=&\left(\frac{r-r_{H}}{r}-\frac{4a^{2}\sin^{2}\theta}{r^{4}}\right)
~u^{t}+\frac{2a\sin^{2}\theta}{r}~u^{\phi},\nn\\
L&=&-\frac{2a\sin^{2}\theta}{r}~u^{t}+r^{2}\sin^{2}\theta~u^{\phi}.
\label{eandl2} \eea

On the equatorial plane, we can readily express $u^{t}$ and
$u^{\phi}$ in terms of $E$ and $L$ as in (\ref{tphidot}) to yield
the radial equation for the particle on the equatorial plane
(\ref{eomr}) with the effective potential \beq
V=-\frac{r_{H}}{2r}\kappa+\frac{L^{2}}{2r^{2}}+\frac{1}{2}\kappa-\frac{L}{r^{3}}
\left[\left(\frac{r_{H}}{2}+\frac{2a^{2}}{r^{3}}\right)L-2aE\right].
\label{effpot2}\eeq  Here note that the last terms associated with
$-L/r^{3}$ dominate over the centrifugal barrier term at small
$r$. In the vanishing $a$ limit (or equally in the $\Omega=0$ case
in (\ref{4metric1})), we can easily see that the effective
potential $V$ in (\ref{effpot2}) reduces to the Schwarzschild
black hole case.

For the null geodesics with $\kappa=0$, we find the only maximum
of the effective potential occurring at \beq
r\ge\frac{3}{2}r_{H}-\frac{6aE}{L}.\eeq  The effective potentials
$V(x,y)$ for the null geodesics of the particles with the total
energy per unit rest mass $E=1$ and angular momentum per unit rest
mass $L=2r_{H}$ are shown in the left graph in Fig. 2. Here $x$
and $y$ represent the dimensionless variables $x=r/r_{H}$ and
$y=a/r_{H}^{2}$. Here observe that along the $y=x^{3}/4$ curve we
have maximal potential values to yield maximal effective potential
values.

\begin{figure}
\begin{center}
\includegraphics[width=10cm]{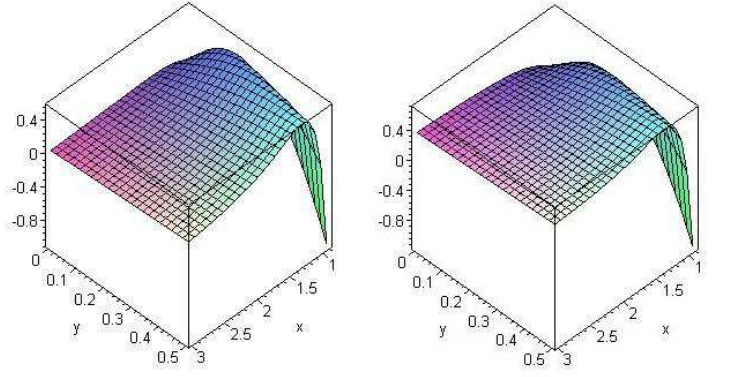}\\
\end{center}
\vskip -.5cm \caption[fig2] {Effective potentials $V(x,y)$ with
$x=r/r_{H}$ and $y=a/r_{H}^{2}$ for null and timelike geodesics of
particles with $E=1$ and $L=2r_{H}$.} \label{fig2}
\end{figure}

For the timelike geodesics with the effective potential
(\ref{effpot2}) with $\kappa=1$, the condition (\ref{v0}) yields
the radius $r_{s}$ of the stable bound orbits which satisfy on the
equatorial plane \beq
r_{H}r_{s}^{5}-2L^{2}r_{s}^{4}+3(L^{2}r_{H}-4aLE)r_{s}^{3}+24a^{2}L^{2}=0.
\eeq   The energy per unit mass of the particle in the circular
orbit of the radius $r=r_{s}$ is found to have the energy lower
bound for $r_{s}^{4}\ge 12a^{2}$ \beq
E_{s}\ge\left(1-\frac{r_{H}}{r_{s}}+\frac{r_{H}(r_{s}^{4}-r_{H}r_{s}^{3}-4a^{2})}
{r_{s}(2r_{s}^{4}-2r_{H}r_{s}^{3}-24a^{2})}\right)^{1/2}.\eeq  The
effective potentials $V(x,y)$ for the null and timelike geodesics
of the particles with the total energy per unit rest mass $E=1$
and angular momentum per unit rest mass $L=2r_{H}$ are shown in
the right graph in Fig. 2 with $x$ and $y$ being the dimensionless
variables $x=r/r_{H}$ and $y=a/r_{H}^{2}$. As in the null geodesic
case, along the $y=x^{3}/4$ curve we have maximal potential values
to yield maximal effective potential values.

The fundamental equations of relativistic fluid dynamics can be
obtained from the conservation of particle number and
energy-momentum fluxes.  In order to derive an equation for the
conservation of particle numbers one can use the particle flux
four vector $nu^{a}$~\cite{landau} \beq
\na_{a}(nu^{a})=\frac{1}{\sqrt{-g}}\na_{a}(\sqrt{-g}~nu^{a})=0,
\label{conteq} \eeq where $n$ is the proper number density of
particles measured in the rest frame of the fluid and $\na_{a}$ is
the covariant derivative in the rotating Schwarzschild curved
manifold and $g={\rm det}~g_{ab}$.  For steady state axisymmetric
flow, the conservation of energy-momentum fluxes is similarly
described by the Einstein equation~\cite{shapiro} \beq
\na_{b}T_{a}^{b}=\frac{1}{\sqrt{-g}}\na_b(\sqrt{-g}~T_{a}^{b})=0,
\label{eineq} \eeq where the stress-energy tensor $T^{ab}$ for
perfect fluid is given by \beq T^{ab}=\rho
u^{a}u^{b}+(g^{ab}+u^{a}u^{b})P\eeq with $\rho$ and $P$ being the
proper internal energy density, including the rest mass energy,
and the isotropic gas pressure, respectively.  The Einstein
equation (\ref{eineq}) can be rewritten in another covariant form
\beq
u_{a}\na_{b}((P+\rho)u^{b})+(P+\rho)u^{b}\na_{b}u_{a}+\na_{a}P=0.
\label{eineq2} \eeq Multiplying (\ref{eineq2}) by $u^{a}$ we can
project it on the direction of the four velocity to obtain \beq
nu^{a}\left(\na_{a}\left(\frac{P+\rho}{n}\right)-\frac{1}{n}\na_{a}P\right)=0,
\label{nua}\eeq where the continuity equation (\ref{conteq}) has
been used.  The radial component of (\ref{nua}) yields \beq
\frac{d\rho}{dr}-\frac{P+\rho}{n}\frac{dn}{dr}=\frac{\Lambda-\Gamma}{u^{r}}.
\label{rrho} \eeq Here the energy loss $\Lambda$ and the energy
gain $\Gamma$ are introduced to set the decrease in the entropy of
inflowing gas equal to the difference $\Lambda-\Gamma$. Moreover,
using the projection operator $g_{ab}+u_{a}u_{b}$ in the equation
(\ref{eineq2}) we can obtain the general relativistic Euler
equation on the direction perpendicular to the four velocity \beq
(P+\rho)u^{b}\na_{b}u_{a}+(g_{ab}+u_{a}u_{b})\na^{b}P=0.
\label{euler} \eeq After some algebra, from (\ref{euler}) we
obtain the radial component of the Euler equation for the steady
state axisymmetric fluid \beq
\frac{d}{dr}(u^{r}u^{r})+\frac{r_{H}}{r^{2}}
+\frac{2}{P+\rho}\left(u^{r}u^{r}+1-\frac{r_{H}}{r}\right)\frac{dP}{dr}=0.
\label{reuler}\eeq Here it is interesting to see that the results
(\ref{rrho}) and (\ref{reuler}) hold also in the rotating
Schwarzschild black hole with the constant $\Omega$ and even in
the static Schwarzschild black hole.

Next, we consider the global embeddings of the rotating
Schwarzschild manifold with the angular velocity
$\Omega=2a/r^{3}$. After tedious algebra, for the rotating
Schwarzschild black hole in the region $r\ge r_{H}$ we can obtain
the (8+6) GEMS structure
\begin{widetext}
\bea ds^{2}&=&-(dz^{0})^{2}+(dz^{1})^{2}+(dz^{2})^{2}+(dz^{3})^{2}
+(dz^{4})^{2}-(dz^{5})^{2}-(dz^{6})^{2}\nn\\
&&-(dz^{7})^{2}+(dz^{8})^{2}+(dz^{9})^{2}+(dz^{10})^{2}-(dz^{11})^{2}
-(dz^{12})^{2}+(dz^{13})^{2} \label{ds2} \eea
\end{widetext}
with the coordinate transformations \bea
z^{0}&=&k_{H}^{-1}\left(\frac{r-r_{H}}{r}\right)^{1/2}\sinh k_{H}t,\nn\\
z^{1}&=&k_{H}^{-1}\left(\frac{r-r_{H}}{r}\right)^{1/2}\cosh k_{H}t,\nn\\
z^{2}&=&\left(\frac{r^{3}+2a}{r}\right)^{1/2}\sin\theta\cos\phi,\nn\\
z^{3}&=&\left(\frac{r^{3}+2a}{r}\right)^{1/2}\sin\theta\sin\phi,\nn\\
z^{4}&=&\left(\frac{r^{3}+2a}{r}\right)^{1/2}\cos\theta,\nn\\
z^{5}&=&\left(\frac{2a}{r}\right)^{1/2}\sin\theta\cos(\phi+t),\nn\\
z^{6}&=&\left(\frac{2a}{r}\right)^{1/2}\sin\theta\sin(\phi+t),\nn\\
z^{7}&=&\left(\frac{2a}{r}\right)^{1/2}\cos\theta,\nn\\
z^{8}&=&\left(\frac{2a(r^{3}+2a)}{r^{4}}\right)^{1/2}\sin\theta\cos t,\nn\\
z^{9}&=&\left(\frac{2a(r^{3}+2a)}{r^{4}}\right)^{1/2}\sin\theta\sin t,\nn\\
z^{10}&=&\left(\frac{4a(r^{3}+2a)}{r^{4}}\right)^{1/2}\cos\theta,\nn\\
z^{11}&=&\left(\frac{2a(r^{3}+2a)}{r^{4}}\right)^{1/2}\sin\theta,\nn\\
z^{12}&=&\left(\frac{r^{3}+2a}{r}\right)^{1/2},\nn\\
z^{13}&=&\int
dr\left(\frac{2r_{H}(r^{2}+rr_{H}+r_{H}^{2})+2r^{3}+a}{2r^{3}}\right)^{1/2}\nn\\
&\equiv& g(r), \label{gems14} \eea where $k_{H}$ is given by
(\ref{sg}).  Note that in the limit $a\rightarrow 0$ the embedding
(\ref{gems14}) reduces to the Schwarzschild case.  In fact,
$(z^{0},z^{1},z^{2},z^{3},z^{4})$ go to those in the Schwarzschild
black hole (\ref{sch6}) and
$(z^{5},z^{6},z^{7},z^{8},z^{9},z^{10},z^{11})$ disappear.
Moreover, we find
\bea &&-(dz^{12})^{2}+(dz^{13})^{2}
=\nn\\&&-\frac{(r^{3}-a)^{2}}{r^{3}(r^{3}+2a)}dr^{2}
+\frac{2r_{H}(r^{2}+rr_{H}+r_{H}^{2})+2r^{3}+a}{2r^{3}}dr^{2},\nn\\
\eea
which , in the vanishing $a$ limit, becomes $(df)^{2}$ in
(\ref{gems6}) so that $(z^{12},z^{13})$ merges into $z^{5}$ in the
Schwarzschild black hole (\ref{sch6}). When the black hole rotates
sufficiently fast, $g_{tt}$ becomes negative as in the Kerr black
hole to yield the ergosphere where \beq r_{H}<r<2a\sin\theta.
\label{ergo} \eeq Here note that the above embedding
(\ref{gems14}) covers without any singularities the whole patch on
$r>r_{H}$ inert to the presence of the ergosphere.

Next, in order to investigate the region $r\le r_{H}$ we rewrite
the rotating Schwarzschild black hole 4-metric (\ref{4metric2}) as
\beq ds^{2}=\bar{N}^{2}dt^{2}-\bar{N}^{-2}dr^2+r^{2}d\theta^{2}
+r^{2}\sin^{2}\theta\left(d\phi-\frac{2a}{r^{3}} dt\right)^{2},
\eeq in terms of the positive definite lapse function (\ref{nbar})
inside the event horizon $r_{H}$ to yield the (8+6) GEMS structure
(\ref{ds2}) with the coordinate transformation \bea
z^{0}&=&k_{H}^{-1}\left(\frac{r_{H}-r}{r}\right)^{1/2}\cosh k_{H}t,\nn\\
z^{1}&=&k_{H}^{-1}\left(\frac{r_{H}-r}{r}\right)^{1/2}\sinh k_{H}t,\nn\\
z^{13}&=&g(r), \label{gems142} \eea
with $(z^{2},z^{3},z^{4},z^{5},z^{6},z^{7},z^{8},z^{9},z^{10},z^{11},z^{12})$ in (\ref{gems14}).

\section{Conclusions}
\setcounter{equation}{0}
\renewcommand{\theequation}{\arabic{section}.\arabic{equation}}

In conclusion, taking an ansatz for a rotating Schwarzschild black
hole analogous to the rotating Schwarzschild wormhole~\cite{teo98}
we have investigated hydrodynamic properties of the perfect fluid
spiraling inward toward the black holes along a conical surface.
Here we have exploited the fact that the coordinates $t$ and
$\phi$ are cyclic in the rotating Schwarzschild metric to find the
timelike Killing field and the axial Killing field, to which we
could obtain the conserved energy and the angular momentum per
unit rest mass for geodesics.

On the equatorial plane of the rotating Schwarzschild black hole,
we have derived the radial equations of motion with the effective
potential.  We have also performed numerical analysis of the
effective potentials of particles on the rotating Schwarzschild
manifolds in terms of angular velocity, total energy and angular
momentum per unit rest mass.  Finally, we have studied the
rotating Schwarzschild black hole manifolds to construct (8+6)
higher dimensional global embeddings inside and outside the event
horizons.

\acknowledgments
We would like to acknowledge financial support in
part from the Korea Science and Engineering Foundation Grant
R01-2000-00015.

\end{document}